\documentclass[showpacs,preprint,preprintnumbers,amsmath,amssymb]{revtex4}
\usepackage{amssymb}
\usepackage{graphicx,psfrag}
\usepackage{amsfonts}
\usepackage{amsmath}
\usepackage{dcolumn}
\usepackage{bm}
\usepackage{epsfig,subfigure}

\begin{document}

\title{Particle collisions in the lower dimensional rotating black hole space-time with the cosmological constant}

\author{Jie Yang$^1$}
\thanks{E-mail: yangjiev@lzu.edu.cn, corresponding author}

\author{Yun-Liang Li$^1$}
\thanks{E-mail: liyunl09@lzu.edu.cn}

\author{Yang Li$^2$}
\thanks{E-mail: liyang12@mail.bnu.edu.cn}

\author{Shao-Wen Wei$^1$}
\thanks{E-mail: weishw@lzu.edu.cn}

\author{Yu-Xiao Liu$^1$}
\thanks{E-mail: liuyx@lzu.edu.cn}
\affiliation{$^1$Institute of Theoretical Physics, Lanzhou University, Lanzhou 730000, China}
\affiliation{$^2$Department of Physics, Beijing Normal University, Beijing 100875, China}

\begin{abstract}
In this paper, we study the effect of ultra-high energy collisions of two particles with different energies near the horizon of a 2+1 dimensional BTZ black hole (BSW effect). We find that the particle with the critical angular momentum could exist inside the outer horizon of the BTZ black hole regardless of the particle energy. Therefore, for the non-extremal BTZ black hole, the BSW process is possible on the inner horizon with the fine tuning of parameters which are characterized by the motion of particle. While for the extremal BTZ black hole, the particle with the critical angular momentum could only exist on the degenerate horizon, and the BSW process could also happen there.
\end{abstract}


\pacs{97.60.Lf, 04.70.-s}

\maketitle

\section{Introduction}

In the recent paper \cite{Banados1}, Ban\~{a}dos, Silk and West proposed a mechanism (BSW process) that two particles may collide on the horizon of an extremal Kerr black hole with ultra-high center-of-mass (CM) energy. Although it was pointed out in Refs. \cite{Berti} and \cite{Jacobson} that the collision
in fact takes an infinite proper time. Moreover, there are astrophysical limitations preventing a Kerr black hole to be an extreme one, and the gravitational radiation and backreaction effects should also be included in this process. Due to the potential interest in
exploring ultra-high energy physics, the BSW process has been studied extensively in other kinds of black holes or naked singularities \cite{Grib1,Lake,Grib2,Wei1,Grib3,Zaslavskii1,Wei2,Mao,Harada1,Harada2,Grib4,Banados2,Patil,Zaslavskii2,Zaslavskii3,WuPRD,WuPRD2,Kimura,Hackmann,Fernando,Sharif,
Stuchlik,Ding,BejgerPRL}. To achieve ultra-high CM energy under the astrophysical limitation of maximal spin, the multiple scattering was taken into account in the nonextremal Kerr black hole \cite{Grib2,Grib4}. Another more direct application is to consider different extreme rotating black holes, such as Kerr-Newman black holes and the Sen black hole \cite{Wei1,Wei2}. On the other hand, a general explanation of this BSW process was tried to give for a rotating black hole \cite{Zaslavskii2} and for other black holes \cite{Zaslavskii3,WuPRD}. Some efforts had also been made to draw some implications concerning the effects of gravity generated by colliding particles in \cite{Kimura}.

However, all of the works mentioned above have been focused on the black holes embedded in the asymptotically flat space-time without cosmological constant. In our previous work \cite{LiyangCQG}, we had considered the BSW process in the background of Kerr-(anti-) de Sitter black hole with nonzero cosmological constant and had found that the cosmological constant has an important effect on the result. For the case of general Kerr-de Sitter black hole (with positive cosmological constant), the collision of two particles can take place on the outer horizon of the non-extremal black hole and the CM energy of collision can blow up arbitrarily if one of the colliding particles has the critical angular momentum. In the present paper, we extend the investigation of the BSW process to the background of a 2+1 dimensional BTZ black hole \cite{BTZ1992}, and our motivation is to examine whether the BSW effect remains valid in the lower dimensional case. Actually, in Ref. \cite{Lake}, Lake had pointed the divergence of the CM energy of particle collision on the inner horizon of the BTZ black hole, but the process was not discussed in detail. In this paper, we study this process in the BTZ black hole with circumstances.

This paper is organized as follows. In Sec. \ref{SecBHs}, we give a brief review of the BTZ black hole. In Sec. \ref{SecCMEnergy}, we study the CM energy of the particle collision on the horizon, and derive the critical angular momentum to blow up the CM energy. In Sec. \ref{SecRadialMotion}, we investigate the radial motion of colliding particles with the critical angular momentum in detail. The extremal and non-extremal cases are examined respectively. The conclusion is given in the last section.

\section{The 2+1 dimensional BTZ black hole}
\label{SecBHs}

In this section we would like to study the horizon structure of the 2+1 dimensional BTZ black hole.
The metric of the BTZ black hole is usually written as~\cite{BTZ1992} (with units $c=G=1$)
\begin{eqnarray}
ds^2 =-N^2_r dt^2 + N^{-2}_r dr^2 + r^2(N_\phi dt + d\phi)^2 \label{metric}
\end{eqnarray}
with
\begin{eqnarray}
N^2_r(r)&=&-M+\frac{r^2}{l^2}+\frac{J^2}{4r^2},\\
N_\phi(r) &=& -\frac{J}{2r^2},
\end{eqnarray}
where $M$ and $J$ are the mass and spin angular momentum of the black hole, respectively. And $l^2$ is related to the cosmological constant $\Lambda$ by $l^{-2} = -\Lambda$.

The horizons can be solved from $N_r|_{r=r_\text{h}} = 0$, and they are given by
\begin{eqnarray}
r_\pm =\sqrt{\frac{l}{2} (lM \pm \sqrt{l^2M^2 - J^2})}.
\end{eqnarray}
Here, $r_+$ is the outer horizon and $r_-$ is the inner horizon. The existence of the horizon requires
\begin{eqnarray}
|J|\leq Ml .
\end{eqnarray}
The horizon of the extremal black hole (corresponding to $|J|= Ml$) is read as
\begin{eqnarray}
r_e = \sqrt{\frac{M}{2}}l.
\end{eqnarray}

\section{The center-of-mass energy for the on-horizon collision}
\label{SecCMEnergy}

To investigate the CM energy of the collision on the horizon of the BTZ black hole, we have to derive the 2+1 dimensional ``4-velocity'' of the colliding particle in the background of the 2+1 dimensional BTZ black hole.

The generalized momentum $P_\mu$ is
\begin{equation}
P_\mu=g_{\mu \nu}\dot{x}^\nu,
\end{equation}
where the dot denotes the derivative with respect to the affine parameter $\lambda$ and $\mu, \nu = t, r, \phi$. Thus, the components $P_t$ and $P_{\phi}$ of the momentum are turned out to be
\begin{eqnarray}
P_t &=& g_{tt} \dot{t} + g_{t \phi} \dot{\phi},\label{4-veloE} \\
P_\phi &=& g_{\phi \phi} \dot{\phi} + g_{t \phi} \dot{t}.\label{4-veloL}
\end{eqnarray}
$P_t$ and $P_\phi$ are constants of motion. In fact, they are correspond to the test particle's energy per unit mass $E$ and the angular momentum parallel to the symmetry axis per unit mass $L$, respectively. And in the following discussion we will just regard these two constants of motion as $-E\equiv P_t$ and $L\equiv P_\phi$ \cite{Hackmann}.

The affine parameter $\lambda$ can be related to the proper time by $\tau = \mu \lambda$, where $\tau$ is given by the normalization condition $-{\mu}^2=g_{\mu \nu}\dot{x}^\mu \dot{x}^\nu$ with $\mu^2 = 1$ for time-like geodesics and $\mu^2=0$ for null geodesics. For a time-like geodesic, the affine parameter can be identified with the proper time, and thus from above equations (\ref{4-veloE}) and (\ref{4-veloL}), we can solve the 2+1 dimensional ``4-velocity'' components $\dot{t}$ and $\dot{\phi}$ (where the dot denotes a derivative with respect to the proper time now) as
\begin{eqnarray}
\frac{dt}{d\tau}&=&\frac{2 {E}-\frac{J L}{r^2}}{2N^2_r}, \label{tt} \\
\frac{d\phi}{d\tau}&=&\frac{J \left(-J L+2 {E} r^2\right)+4 L r^2 N^2_r}{4 r^4 N^2_r}.\label{phit}
\end{eqnarray}
For the remained component $\dot{r} = \frac{dr}{d\tau}$ of the radial motion, we can obtain it from the Hamilton-Jacobi equation of the time-like geodesic
\begin{eqnarray}
\frac{\partial S}{\partial \tau} = -\frac{1}{2}g^{\mu \nu}\frac{\partial S}{\partial x^\mu}\frac{\partial S}{\partial x^\nu}\label{HYE}
\end{eqnarray}
with the ansatz
\begin{eqnarray}
S = \frac{1}{2}\tau - Et + L\phi + S_r(r),
\end{eqnarray}
where $S_r(r)$ is a function of $r$. Inserting the ansatz into (\ref{HYE}), with the help of the metric (\ref{metric}), we get
\begin{eqnarray}
\left(\frac{dS_r(r)}{dr} \right)^2 &=& \frac{J^2 L^2-4 E J L r^2-4 r^2 \left[L^2 N^2_r+\left(-E^2+N^2_r\right) r^2\right]}{4 N^4_r r^4}
.
\end{eqnarray}
On the other hand, we have
\begin{eqnarray}
\frac{dS_r(r)}{dr} = P_r= g_{rr}\dot{r} = \frac{\dot{r}}{N^2_r}.
\end{eqnarray}
Thus we get square of the radial component of the 4-velocity
\begin{eqnarray}
\left(\frac{dr}{d\tau}\right)^2 &=& \frac{K^2-4 r^2 N^2_r\left(L^2+r^2\right)}{4 r^4} , \label{rt}
\end{eqnarray}
where
\begin{eqnarray}
K&=&J L-2 {E} r^2.
\end{eqnarray}
Here we have obtained all nonzero 2+1 dimensional ``4-velocity'' components for the geodesic equation. Next we would like to study the CM energy of the two-particle collision in the background of the BTZ black hole. Here we consider more general case that the two colliding particles have different energies $E_1$, $E_2$, and different angular momenta per unit mass $L_1$, $L_2$. For simplicity, the particles under consideration have the same rest mass $m_0$. We can compute the CM energy $E_{\text{CM}}$ of this two-particle collision by using
\begin{eqnarray}
E_{\text{CM}} = \sqrt{2} m_0 \sqrt{1-g_{\mu \nu}u^{\mu}_1u^{\nu}_2},
\end{eqnarray}
where $u^{\mu}_1, u^{\nu}_2$ are the ``4-velocity'' vector of the two particles ($u=(\dot{t},\, \dot{r},\, \dot{\phi})$). With the help of Eqs. (\ref{tt}), (\ref{phit}), and (\ref{rt}), we obtain the CM energy
\begin{eqnarray}
 \frac{E_{\text{CM}}^2}{2m_0^2}
  &=& \frac{1}{4 r^4 N^2_r}\Big[\left(J L_1 -2 E_1 r^2\right) \left(J L_2 -2 E_2 r^2\right) \nonumber \\
  &&~~~~~~~~~~+4 r^2 N^2_r\left(- L_1 L_2+r^2\right)  - H_1 H_2\Big], \label{Ecm}
\end{eqnarray}
where
\begin{eqnarray}
H_i = \sqrt{\left(J L_i - 2 E_i r^2\right)^2-4 r^2 N^2_r\left(L_i^2+r^2\right) }
~~(i=1, 2).
\end{eqnarray}
For simplicity, we can rescale the CM energy as $\bar{E}_{\text{CM}}^2 \equiv \frac{1}{2m_0^2}E_{\text{CM}}^2$. We would like to study $\bar{E}_{\text{CM}}^2$ for the case that the particles collide on the black hole's horizon, which means $N_r=0$. The denominator of the expression on the right hand of (\ref{Ecm}) is zero, and the numerator of it is
\begin{eqnarray}
&&K_1K_2-\sqrt{K_1^2}\sqrt{K_2^2},\\
&&K_i=K|_{E=E_i, L=L_i},\,\,  i=1, 2.
\end{eqnarray}
When $K_1K_2\geq0$, the numerator will be zero and the value of $\bar{E}_{\text{CM}}^2$ on the horizon will be undetermined; but when $K_1K_2<0$, the numerator will be negative finite value and $\bar{E}_{\text{CM}}^2$ on the horizon will be negative infinity. So it should have $K_1K_2\geq0$, and for the CM energy on the horizon, we have to compute the limiting value of Eq. (\ref{Ecm}) as $r \rightarrow r_\text{h}$, where $r_\text{h}$ is the horizon of the black hole.

After some calculations, we get the limiting value of Eq. (\ref{Ecm}):
\begin{eqnarray}
\bar{E}_{\text{CM}}^2(r \rightarrow r_\text{h})&=&2+\frac{(L_1-L_2)^2-l^2(E_1-E_2)^2 -2 (L_1-L_2) (L_{\text{C}1}-L_{\text{C}2})}{2 (L_1-L_{\text{C}1}) (L_2-L_{\text{C}2})}\nonumber \\
&&+\frac{l \left[(E_2 L_1-E_1 L_2)^2+M l^2 (E_1-E_2)^2 \right] \left(l M+\sqrt{l^2 M^2-J^2}\right)}{J^2 (L_1-L_{\text{C}1}) (L_2-L_{\text{C}2})} ,
\end{eqnarray}
which can also be rewritten as
\begin{eqnarray}
\bar{E}_{\text{CM}}^2(r \rightarrow r_\text{h})=2+\frac{A }{2 K_1 K_2},
\end{eqnarray}
where
\begin{eqnarray}
A&=&J^2 \Big[(L_1-L_2)^2-(E_1-E_2)^2 l^2-2 (L_1-L_2) (L_{\text{C}1}-L_{\text{C}2})\Big]\nonumber \\
&&+ 2 l\Big[(E_2 L_1-E_1 L_2)^2+(E_1-E_2)^2 l^2 M\Big] \Big(l M+\sqrt{l^2 M^2-J^2}\Big).
\end{eqnarray}
So it can be seen that when $K_i=0$, the CM energy on the horizon will blow up.
Solving $K_i=0$, we get the critical angular momentum
\begin{eqnarray}
L_{\text{C}i}=\frac{2r_\text{h}^2E_i}{J }=\frac{E_i l \left(l M+\sqrt{l^2 M^2-J^2}\right)}{J}\label{Lc}, \,\, i=1,2.
\end{eqnarray}
It is easy to prove that when $K_1=0$ and $K_2=0$, the CM energy is finite. So in order to obtain an arbitrarily high CM energy, one and only one of the colliding particles should have the critical angular momentum. For the extremal BTZ black hole $J=lM$, the $\bar{E}_{\text{CM}}^2$ on the extremal horizon is
\begin{eqnarray}
\bar{E}_{\text{CM}}^2(r \rightarrow r_e)=2 + \frac{{M}\big[(L_1-E_1 l)-(L_2 -E_2 l)\big]^2+{2(E_2 L_1-E_1 L_2)^2}}{2 {M}(L_1-E_1 l) (L_2-E_2 l)}.
\end{eqnarray}
Obviously, when one particle has the critical angular momentum $L_{\text{C}1}=E_{\text{1}} l$ (or $L_{\text{C}2}=E_{\text{2}} l$), and the other does not, the CM energy on the extremal horizon could be infinite.

From above derivation, it seems that the CM energy could blow up on the horizon. However, in order to get arbitrarily high CM energy on the horizon of the BTZ black hole, the colliding particle with the critical angular momentum must be able to reach the horizon of the black hole. We will investigate this part in the next section.

\section{The radial motion of the particle with the critical angular momentum near the horizon}
\label{SecRadialMotion}

\begin{figure}
\includegraphics[width=80mm,height=60mm]{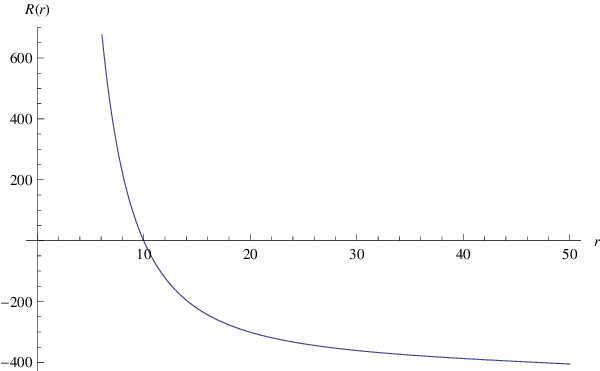}
\includegraphics[width=80mm,height=60mm]{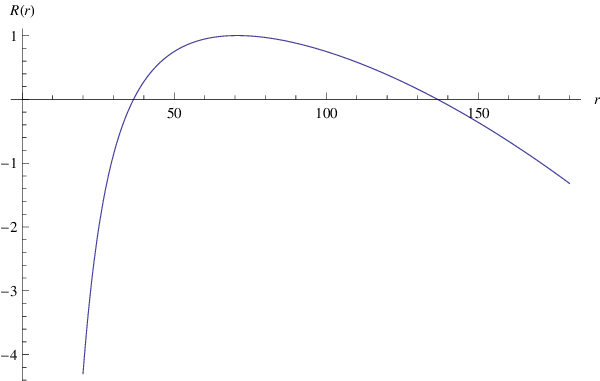}
\caption{Behaviors of $R(r)$. Left for $L^2 M-E J L-\frac{J^2}{4 }>0$ and right $L^2 M-E J L-\frac{J^2}{4 }<0$.}
\label{RrCriLWW}
\end{figure}

In this section, we will study the radial motion of the particle with the critical angular momentum and find the region it can exist. In order for a particle to reach the horizon of the black hole, the square of the radial component of the ``4-velocity'' $\left(\frac{dr}{d\tau}\right)^2$ in Eq. (\ref{rt}) has to be positive in the neighborhood of the black hole's horizon. Obviously, $R(r)|_{L=L_{\text{C}i}}=0$ on the horizon of the BTZ black hole. For a particle with arbitrary energy $E$ and angular momentum $L$, the explicit form of $({dr}/{d\tau})^2$, which is denoted by $R(r)$, reads

\begin{equation}
R(r)\equiv\left(\frac{dr}{d\tau}\right)^2 =E^2-\frac{L^2}{l^2}+M+\frac{1}{r^2}\left(L^2 M-EJL-\frac{J^2}{4 }\right)-\frac{r^2}{l^2}.\label{rr}
\end{equation}
We draw $R(r)$ in Fig.~\ref{RrCriLWW}. It can be seen that, when $
L^2 M-E J L-{J^2}/{4 }>0$, $R(r\rightarrow 0) \rightarrow +\infty$ and $R(r\rightarrow +\infty) \rightarrow -\infty$, so there is only one positive root for $R(r)=0$ and the particle can exist in the region inside of the root. When $L^2 M-E J L-{J^2}/{4 }<0$,
$R(r\rightarrow 0) \rightarrow -\infty$ and $R(r\rightarrow +\infty) \rightarrow -\infty$, there are two positive roots and the particle can exist in the region between the two roots.
The bigger root of $R(r)=0$ is
\begin{eqnarray}
r_2=\frac{\sqrt{l^2\left(E^2+M\right)-L^2 +\sqrt{\left[l^2 M-J l+(L - E l)^2\right] \left[l^2 M+J l+(L + E l)^2\right]}}}{\sqrt{2}}.
\end{eqnarray}
We find that it increases with $E$ and $L$, which means the particle can move to arbitrarily far from black hole's horizon with it's energy and angular momentum increase.

Next,  we will study the radial motion of the particle with the critical angular momentum:
\begin{eqnarray}
R(r)|_{L=L_\text{c}}&=& \frac{W}{ r^2}-\frac{r^2}{l^2}+2E^2+M-\frac{2 E^2 l^2 M^2+2E^2 l M \sqrt{l^2 M^2-J^2}}{J^2}, \label{RrwithcriL}
\end{eqnarray}
where
\begin{eqnarray}
W=\frac{E^2 l\left[2l M\left(l^2 M^2-J^2 \right)+\left(2 l^2 M^2-J^2\right)\sqrt{l^2 M^2-J^2}\;\right]}{J^2 }-\frac{J^2}{4 }.
\end{eqnarray}
By solving $W=0$ we get the critical energy $E_0$:
\begin{eqnarray}
E_0=\frac{J^2}{2 \sqrt{2 l^2 M\left(l^2 M^2- J^2\right)+\left(2 l^3 M^2-J^2l\right) \sqrt{l^2 M^2-J^2}}}.
\end{eqnarray}
When $E>E_0$, $R(r)=0$ has one root
\begin{eqnarray}
r_0&=&\frac{1}{\sqrt{2} J}\Big\{l^2 \big[J^2 M+2 E^2 (J^2-l M (l M+\sqrt{l^2 M^2 - J^2}))\big]\nonumber\\
&&+l\big[(l^2 M^2-J^2) [J^4+8 E^4 l^3 M (l M+\sqrt{l^2 M^2-J^2})\nonumber\\
&&+4 E^2 J^2 l (l M - E^2 l+\sqrt{l^2 M^2 - J^2})]\big]^\frac{1}{2}\Big\}^\frac{1}{2}
\end{eqnarray}
and the particle with the critical angular momentum can exist inside of it. When $E<E_0$, $R(r)=0$ has two roots
\begin{eqnarray}
r_{0+}&=&\frac{1}{\sqrt{2} J}\Big\{l^2 \big[J^2 M+2 E^2 (J^2-l M (l M+\sqrt{+l^2 M^2 - J^2}))\big]\nonumber\\
&&+l\big[(l^2 M^2-J^2) (J^4+8 E^4 l^3 M (l M+\sqrt{l^2 M^2-J^2})\nonumber\\
&&+4 E^2 J^2 l (l M - E^2 l+\sqrt{l^2 M^2 - J^2}))\big]^\frac{1}{2}\Big\}^\frac{1}{2},
\end{eqnarray}
\begin{eqnarray}
r_{0-}&=&\frac{1}{\sqrt{2} J}\Big\{l^2 \big[J^2 M+2 E^2 (J^2-l M (l M+\sqrt{+l^2 M^2 - J^2}))\big]\nonumber\\
&&-l\big[(l^2 M^2-J^2) (J^4+8 E^4 l^3 M (l M+\sqrt{l^2 M^2-J^2})\nonumber\\
&&+4 E^2 J^2 l (l M - E^2 l+\sqrt{l^2 M^2 - J^2}))\big]^\frac{1}{2}\Big\}^\frac{1}{2},
\end{eqnarray}
and the particle with the critical angular momentum can exist between them.
The above discussion only concerns about the square of the radial component of the ``4-velocity''. To find whether the particle with the critical angular momentum can reach the horizon of the BTZ black hole, we should investigate the roots of $R(r)=0$ and the horizons of the black hole. The non-extremal and extremal cases will be considered in following.

\subsection{Non-extremal BTZ black hole}
For the non-extremal BTZ black hole case, we can prove that the solution (for $E>E_0$ case) or the bigger solution (for $E<E_0$ case) of $R(r)=0$ is just the outer horizon of black hole
\begin{eqnarray}
r_0=r_+ = \sqrt{\frac{l}{2} (l M + \sqrt{l^2M^2 - J^2})}.
\end{eqnarray}
That means the particle with the critical angular momentum can exist inside the outer horizon of the non extremal BTZ black hole.
\begin{figure}
\includegraphics[width=80mm,height=60mm]{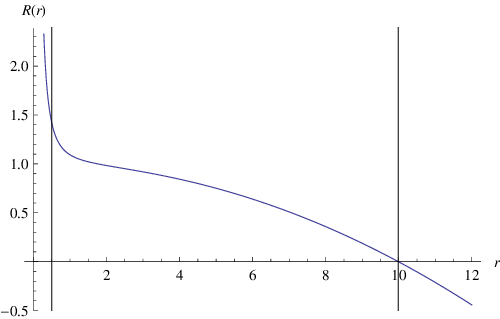}
\includegraphics[width=80mm,height=60mm]{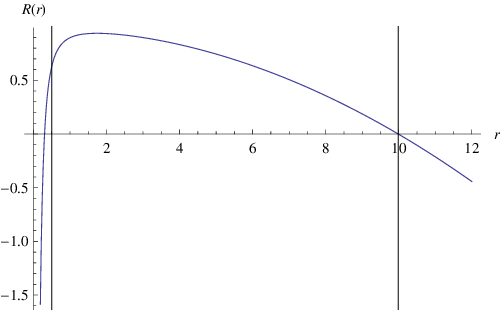}
\caption{The variation of $R(r)$ vs radius $r$ for the case of the non-extremal BTZ black hole ($l=10$, $M=1$, $J=1$), $E>E_0$ (left, $E=0.003$), $E<E_0$ (right, $E=0.002$). The
vertical lines denote the locations of the outer and inner horizons.}
\label{RrCriticalLWithhorizon}
\end{figure}

\subsection{Extremal BTZ black hole}

For the extremal BTZ black hole case,
$R(r)$ for particle with the critical angular momentum becomes very simple
\begin{eqnarray}
R(r)=M-\frac{r^2}{l^2}-\frac{J^2}{4r^2 }.
\end{eqnarray}
We solve $R(r)=0$ and get
\begin{eqnarray}
r_0= \sqrt{\frac{M}{2}}l.
\end{eqnarray}
It is just the degenerated horizon of the extremal black hole.
\begin{figure}
\includegraphics[width=100mm]{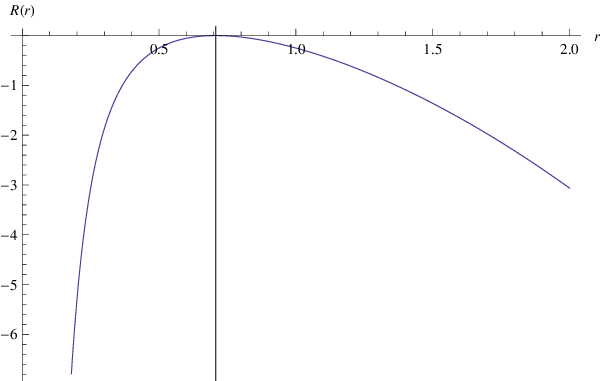}
\caption{The variation of $R(r)$ vs radius r for for the case of the extremal BTZ black hole, ($l=1$, $M=1$, $J=1$), The
vertical line denotes the locations of the degenerated horizon.}
\label{RrCriLExtrBHWithhorizon}
\end{figure}

The behaviors of $R(r)$ for the particle with the critical angular momentum are plotted in fig. \ref{RrCriticalLWithhorizon} for the non-extremal black hole and fig. \ref{RrCriLExtrBHWithhorizon} for the extremal black hole. for the non-extremal black hole, we find the particle with the critical angular momentum can exist inside the outer horizon. So particle collision on the inner horizon could produce unlimited CM energy. For the extremal black hole, the particle with the critical angular momentum could only exist on the degenerated horizon, so if such particle exists, then unlimited CM energy will be approached.

\section{Conclusion}\label{Conclusion}
In this work, we have analyzed the possibility that the 2+1 dimensional BTZ black holes can serve as particle accelerator. We first calculate the CM energy for the two-particle collision. In order to obtain a unlimited CM energy, one of the particles should have the critical angular momentum. Next, we study the radial motion for the particle with the critical angular momentum. For the extremal BTZ black hole, particles with critical angular momentum can only exist on the outer horizon of the BTZ black hole. So if such particle exist, then unlimited CM energy will be approached. For the non-extremal BTZ black hole, particles can collide on the inner horizon with arbitrarily high CM energy.

\section*{Acknowledgements}
This work was supported by the National Natural Science Foundation of China (Grants No.
11205074 and No. 11375075), and the Fundamental Research Funds for the Central Universities (Grants
No. lzujbky-2013-18 and No. lzujbky-2013-21).

\end{document}